\documentclass[sigconf]{acmart}
\usepackage{xspace}
\usepackage{lipsum}
\usepackage[svgnames]{xcolor}
\usepackage{calc}
\usepackage{pgfmath}
\usepackage[bb=boondox,bbscaled=.95,cal=boondoxo]{mathalfa}
\usepackage{graphicx}
\usepackage{booktabs}
\usepackage{amsfonts}
\usepackage{amssymb}  
\usepackage{float} 
\usepackage{array}
\usepackage{caption}
\usepackage{xcolor}
\usepackage{subcaption}
\usepackage{makecell}
\usepackage{amssymb, amsmath}
\usepackage[absolute,overlay]{textpos}
\usepackage[table]{xcolor}
\usepackage{multirow}
\usepackage{caption}
\usepackage{array}
\usepackage{tikz}
\usepackage{bbding}
\usepackage{pifont}
\usepackage{wasysym}
\usepackage{cleveref}
\usepackage{fontawesome}
\usepackage{utfsym}
\usepackage{tabularx}
\usepackage{xcolor}
\usepackage{colortbl}
\usepackage{xurl}
\usepackage{algorithm}
\usepackage{algorithmic}

\newcommand{\mypara}[1]{\smallskip\noindent{\bf {#1}.} \xspace}
\newcommand{\ib}{IV-Bridge\xspace}
\newcommand{\bench}{FakeI2V-Bench\xspace}
\newcommand{\Cfill}{\CIRCLE}
\newcommand{\copen}{\Circle}
\newcommand{\lc}{\LEFTcircle}

\newcommand{\gammaa}{0.5}
\newcommand{\Tcap}{16000} 
\newcommand{\Tmax}{16000} 
\newcommand{\timebargreen}[2]{
  \begin{tikzpicture}[baseline=(txt.base)]
    \def\W{1.2cm}
    \def\BW{1.25cm} 
    \def\H{1.6ex}
    \path (0,0) rectangle (\W,\H);
    \pgfmathsetmacro{\t}{min(#1,\Tcap)/\Tmax}
    \pgfmathsetmacro{\p}{pow(\t,\gammaa)}
    \pgfmathsetmacro{\p}{min(max(\p,0),1)}
    \pgfmathsetlengthmacro{\barL}{\BW*\p}
    \fill[Tomato!50] ({\W-\barL},0) rectangle (\W,\H);
    \node[anchor=east, inner sep=0pt] (txt) at (\W,\H/2) {#2};
  \end{tikzpicture}%
}
\newcommand{\Tcapp}{550}
\newcommand{\Tmaxp}{550} 
\newcommand{\parabarblue}[2]{
  \begin{tikzpicture}[baseline=(txt.base)]
    \def\W{1.6cm} 
    \def\BW{1.25cm} 
    \def\H{1.6ex}
    \path (0,0) rectangle (\W,\H);
    \pgfmathsetmacro{\t}{min(#1,\Tcapp)/\Tmaxp}
    \pgfmathsetmacro{\p}{pow(\t,\gammaa)}
    \pgfmathsetmacro{\p}{min(max(\p,0),1)}
    \pgfmathsetlengthmacro{\barL}{\BW*\p}
    \fill[SandyBrown!60] ({\W-\barL},0) rectangle (\W,\H);
    \node[anchor=east, inner sep=0pt] (txt) at (\W,\H/2) {#2};
  \end{tikzpicture}%
}
\definecolor{aored}{rgb}{0.7, 0.11, 0.11}
\definecolor{aogreen}{rgb}{0.55, 0.71, 0.0}
\definecolor{lightgray}{RGB}{230,230,230}
\definecolor{deepred}{RGB}{200, 0, 0}
\definecolor{deepgreen}{RGB}{0, 137, 0}
\newcommand{\yes}{\textcolor{deepgreen}{\ding{51}}\xspace}
\newcommand{\no}{\textcolor{deepred}{\ding{55}}\xspace}
\usepackage[most]{tcolorbox}
\tcbset{width=0.48\textwidth,boxrule=1pt,colback=lightgray,arc=1pt,auto outer arc,left=0pt,right=0pt,top=0pt,bottom=0pt,boxsep=2pt}
\newcolumntype{Y}{>{\centering\arraybackslash}X}

\AtBeginDocument{%
  }
    
\copyrightyear{2026}
\acmYear{2026}
\setcopyright{cc}
\setcctype{by}
\acmConference[KDD 2026] {Proceedings of the 32nd ACM SIGKDD Conference on Knowledge Discovery and Data Mining V.2}{August 9--13, 2026}{Jeju Island, Republic of Korea.}
\acmBooktitle{Proceedings of the 32nd ACM SIGKDD Conference on Knowledge Discovery and Data Mining V.2 (KDD 2026), August 9--13, 2026, Jeju Island, Republic of Korea}
\acmISBN{979-8-4007-2259-2/2026/08}
\acmDOI{10.1145/3770855.3817509}

\begin{document}

\begin{textblock*}{20cm}(0.5cm,0.5cm)
\begin{center}
To Appear in KDD 2026, Jeju, Korea, August 9--13, 2026.
\end{center}
\end{textblock*}

\title{FakeI2V-Bench: Benchmarking the Applicability of Image-level Deepfake Detectors for Deepfake Video Detection}

\author{Pei Li}
\orcid{0009-0001-9002-7930}
\affiliation{%
  \institution{School of Cyber Science and Technology, \\ Shandong University,}
  \city{Qingdao, Shandong}
  \country{China}
}
\email{leepy@mail.sdu.edu.cn}

\author{Sihan Chen}
\orcid{0009-0008-0176-9782}
\affiliation{%
  \institution{School of Cyber Science and Technology,\\ Shandong University,}
  \city{Qingdao, Shandong}
  \country{China}
}
\email{202437060@mail.sdu.edu.cn}

\author{Delong Ran}
\orcid{0009-0003-2630-2680}
\affiliation{%
  \institution{Institute for Network Sciences and Cyberspace, BNRist, \\ Tsinghua University,}
  \city{Beijing}
  \country{China}
}
\email{rdl22@mails.tsinghua.edu.cn}

\author{Tianshuo Cong}
\authornote{Corresponding author}
\orcid{0000-0003-3189-8223}
\affiliation{%
  \institution{School of Cryptologic Science and Engineering, \\ Shandong University,}
  \city{Jinan, Shandong}
  \country{China}
}
\email{tianshuo.cong@sdu.edu.cn}

\begin{abstract}
Recent advances in video generation models have significantly intensified the deepfake threat, yet the current deepfake video detection benchmarks remain underdeveloped. 
In particular, the effectiveness of image-level detectors in the video domain has not been systematically assessed.
To fill this gap, we present \bench, a benchmark for evaluating state-of-the-art video-level deepfake detectors in challenging scenarios, with a particular focus on systematically assessing the performance of image-level deepfake detectors in the video domain.
\bench comprises $97,548$ videos, containing content generated by the latest powerful generation models and covering a broader range of categories. 
Using this dataset, we conduct a systematic evaluation of eight video-level detectors and twelve representative image-level detectors.
Experimental results show that the best-performing image-level detector achieves an $80.16\%$ AUC, slightly outperforming the strongest video-level detector (i.e., $79.99\%$ AUC).
Going beyond benchmarking, we present IV-Bridge, a general framework that enhances the applicability of image-level deepfake detectors to videos. 
IV-Bridge employs a random forest model with statistical features to aggregate frame-level predictions, allowing eleven image-level detectors to surpass state-of-the-art video-level approaches, with the best-performing variant achieving a $93.80\%$ AUC. 
Overall, \bench establishes a rigorous benchmark for deepfake video detection and introduces a novel pathway for extending image-level detectors to the video domain, offering new insights and directions for future research.
\footnote{Code and data are available at: \url{https://github.com/CryptoAILab/FakeI2V-Bench}.}
\end{abstract}




\maketitle

\section{Introduction}

Large generation models are at the forefront of Artificial Intelligence (AI) innovation, with Video Generation Models (VGMs) driving significant transformations in digital content production. 
Leading VGMs such as Sora~\cite{videoworldsimulators2024} can generate videos up to 1080p resolution from input text prompts or even an existing image or video. 
These AI-generated videos, referred to as \textit{deepfake videos}, not only offer 
benefits in domains like intelligent news anchors~\cite{AInews} and cinematic visual effects~\cite{AIfilm}, but also introduce potential societal safety risks, such as misleading judicial evidence verification~\cite{AIlaw} or synthetic pornography~\cite{AIporn}.

In response to the deepfake threat, significant research efforts have been devoted to designing video-level deepfake detectors~\cite{YJDMF21_1,WQZQHCZN22_1,ZJWWH23_1,DNPPMAED24_1,YYT24_1,HYZZZYJHJWH24_1,XXJQLXGX24_1,zheng2025d3}.
While these detectors demonstrate promising performance, they still face several fundamental limitations, including limited generalization~\cite{ZQMTYHXY25_1,HYZZZYJHJWH24_1} and substantial computational overhead~\cite{XXJQLXGX24_1}.
Moreover, as video generation models continue to evolve rapidly, it remains unclear how well existing detectors perform on videos synthesized by the latest generation models, underscoring the urgent need for systematic, comprehensive, and up-to-date evaluation.

In parallel, image-level deepfake detectors~\cite{SORAA20_1,CYSGY23_1,ZZNY23_1,UYJ23_1} have achieved remarkable progress, benefiting from large-scale training data and well-established detection pipelines.
Meanwhile, these image-level detectors have also been preliminarily employed in deepfake video detection under various usage patterns, such as averaging frame-level predictions to form a video-level decision~\cite{f3net} or reporting detection performance independently on each frame~\cite{idbench,deepfakebench}.
This also indicates that while the potential of image-level detectors for identifying deepfake videos has been increasingly recognized, their effective adaptation and systematic evaluation in video-level detection scenarios remain insufficiently explored.

\subsection{Our Work}
Motivated by the above observations, in this paper, we present \bench, a comprehensive benchmark for deepfake video detection that systematically evaluates both video-level detectors and image-level detectors under a unified evaluation framework.

\mypara{\bench}
\bench consists of four modules: 
(i) Evaluation dataset: \bench collects four datasets (introduced in \Cref{tab:5_datasets}) containing a total of $97,548$ videos, which fall into two major categories: facial datasets and general datasets, encompassing diverse VGMs (e.g., Face2Face~\cite{JMMCM16_1}, Gen2~\cite{runway2023gen2}, etc) and diverse content (e.g., face, foods, natural landscape, etc).
Notably, the inclusion of GenVidBench (GVB)~\cite{ZQMTYHXY25_1} dataset, which was newly proposed in 2025, highlights the timeliness of our measurement.
(ii) Video-level detectors: \bench includes eight state-of-the-art video-level deepfake detectors (listed in \Cref{tab:video_level_deepfake_detection_methods}), including both face-focused and general-purpose detectors. 
(iii) Image-level detectors: \bench integrates twelve representative image-level detectors (listed in \Cref{tab:13_image_deepfake_detectors}) alongside their enhanced counterparts, which are developed through our enhancement framework, \ib. 
(iv) Evaluation Metrics: \bench utilizes two metrics, Area Under Curve (AUC) and Average Precision (AP), to provide a thorough comparison of detection capabilities.

\mypara{Research Questions}
Based on \bench, we aim to address the following three key Research Questions (\textbf{RQ}s).

\begin{itemize}
\item \textbf{RQ1}: What is the detection capability of current emerging video-level deepfake detectors in complex scenarios?
\item \textbf{RQ2}: What is the performance of naive image-level deepfake detectors on video data? 
\item \textbf{RQ3}: How can image-level detectors be enhanced for higher deepfake video detection performance?
\end{itemize}

\mypara{Evaluation Results on RQ1}
To address \textbf{RQ1}, in~\Cref{sec:bench_video_level}, we conduct a comprehensive evaluation of eight video-level detectors. 
\Cref{tab:vft_all} presents the detection performance of all video-level detectors on the FakeI2V-Bench dataset. 
While some detectors achieve excellent results on specific datasets (e.g., >90\% AUC), their performance can drop to around 50\% AUC on other datasets, highlighting limited generalization. 
Overall, these results indicate that current video-level detectors still have substantial room for improvement to achieve consistently high performance across diverse datasets.

\mypara{Evaluation Results on RQ2}
In~\Cref{sec:benchmark_naive}, we further evaluate twelve naive image-level detectors to tackle \textbf{RQ2}. 
We first analyze their frame-level performance and compare it with their performance on a conventional fake image dataset (named FakeGenImage). 
\Cref{tab:frame_level_result_auc_ap_main} reveals that the performance of naive image-level detectors drops substantially when applied to video frames. 
However, by applying different frame-to-video aggregation strategies to obtain video-level predictions, the performance can be significantly improved.
Notably, one naive image-level detector has even surpassed SOTA video-level detectors (see~\Cref{tab:vft_all}), demonstrating the potential of image-level models for deepfake video detection.

\mypara{Solutions to RQ3}
To address \textbf{RQ3}, in~\Cref{sec:ivbridge}, we introduce \ib, a framework for enhancing and adapting image-level detectors to video scenarios.  
\ib consists of two stages:  
(1) \textit{Video-Frame Fine-Tuning (VFT)}, which fine-tunes image-level detectors to capture video-specific forgery patterns, and  
(2) \textit{Multi-Mode Aggregation (MMA)}, which aggregates predictions across multiple frame-to-video strategies through a random forest model.  
As shown in~\Cref{tab:vft_all}, \ib significantly improves the detection performance compared to naive image-level models, and eleven of twelve \ib-enhanced detectors (detectors with the ``-IV" suffix) successfully surpass SOTA video-level detectors. 
Moreover, \ib-enhanced detectors exhibit stronger generalization across different video generation models, and the deployment costs are much lighter than video-level detectors.  
These results not only demonstrate the practical effectiveness of \ib, but also provide a promising direction for future research in deepfake video detection by leveraging mature image-level detectors.

\mypara{Core Contributions}
In summary, we make the following contributions:

\begin{itemize}

\item We introduce \bench, a comprehensive benchmark for evaluating video-level and image-level deepfake detectors across a wide range of evaluation datasets.
\item We present an enhancement framework named \ib to adapt image-level detectors for deepfake video detection.
\item Extensive results validate that eleven of twelve \ib-enhanced detectors successfully surpass SOTA video-level detectors.

\end{itemize}

\begin{table}[t]
\centering
\caption{Comparison of deepfake video detection benchmarks, with a special focus on their support for image-level detectors. ``Latest Detector" denotes the release time of the latest detector evaluated. In the Image-level column, \Cfill~indicates systematic evaluation, \lc~denotes limited evaluation, and \copen~denotes no evaluation.}
\label{tab:benchmark_compare}
\setlength{\tabcolsep}{2pt}
\begin{tabular}{lccccc}
\toprule
\multirow{2}{*}{\textbf{Name}} & \multirow{2}{*}{\makecell{\textbf{Latest}\\\textbf{Detector}}} & \multirow{2}{*}{\textbf{Image-level}} & \multicolumn{2}{c}{\textbf{Dataset}} \\ 
\cmidrule(lr){4-5}
                       &    &  & \textbf{Facial} &\textbf{General}\\
\midrule
IDBench~\cite{idbench}         & 2021-03   & \lc & \yes & \no \\
DeepfakeBench~\cite{deepfakebench}   & 2023-05  & \lc   & \yes  & \no \\
GenVidBench~\cite{ZQMTYHXY25_1}   & 2024-05   & \copen  & \no & \yes \\
DeMamba~\cite{HYZZZYJHJWH24_1}          & 2024-05   & \lc  & \no & \yes \\
\midrule
\bench (\textbf{Ours})  & 2025-08   & \Cfill   & \yes & \yes \\
\bottomrule
\end{tabular}
\end{table}

\section{Related Works}
A comparative summary of representative deepfake video detection benchmarks is provided in \Cref{tab:benchmark_compare}, from which several key limitations can be identified.
\textit{(i) Limited content diversity.} 
Most existing benchmarks are constructed around a single generation scenario, such as face-centric manipulations or generic text-to-video and image-to-video synthesis, which limits the evaluation of detector generalization across diverse and complex video content.
\textit{(ii) Outdated detector coverage.} 
Many benchmarks evaluate only outdated detectors.
For instance, the latest detectors evaluated in IDBench~\cite{idbench} were released around five years ago, while most detectors tested in GenVidBench~\cite{ZQMTYHXY25_1} date from 2017 to 2022, with only one method released after 2024.
\textit{(iii) Insufficient evaluation of image-level detectors.} 
Some benchmarks (e.g., GenVidBench~\cite{ZQMTYHXY25_1} ) exclude image-level detectors entirely, while others (e.g., IDBench~\cite{idbench} and DeepfakeBench~\cite{deepfakebench}) report only frame-level results without principled video-level aggregation.
Although DeMamba~\cite{HYZZZYJHJWH24_1}  evaluates two image-level detectors and claims to consolidate frame-level predictions to obtain video-level predictions, it does not specify the exact aggregation method used. 
In contrast, our work places particular emphasis on image-level detectors, systematically evaluating their performance and applicability in the video domain under well-defined aggregation protocols, thereby providing new insights into their practical potential for deepfake video detection.

\begin{figure*}[t]
\centering
\includegraphics[width=0.9\linewidth]{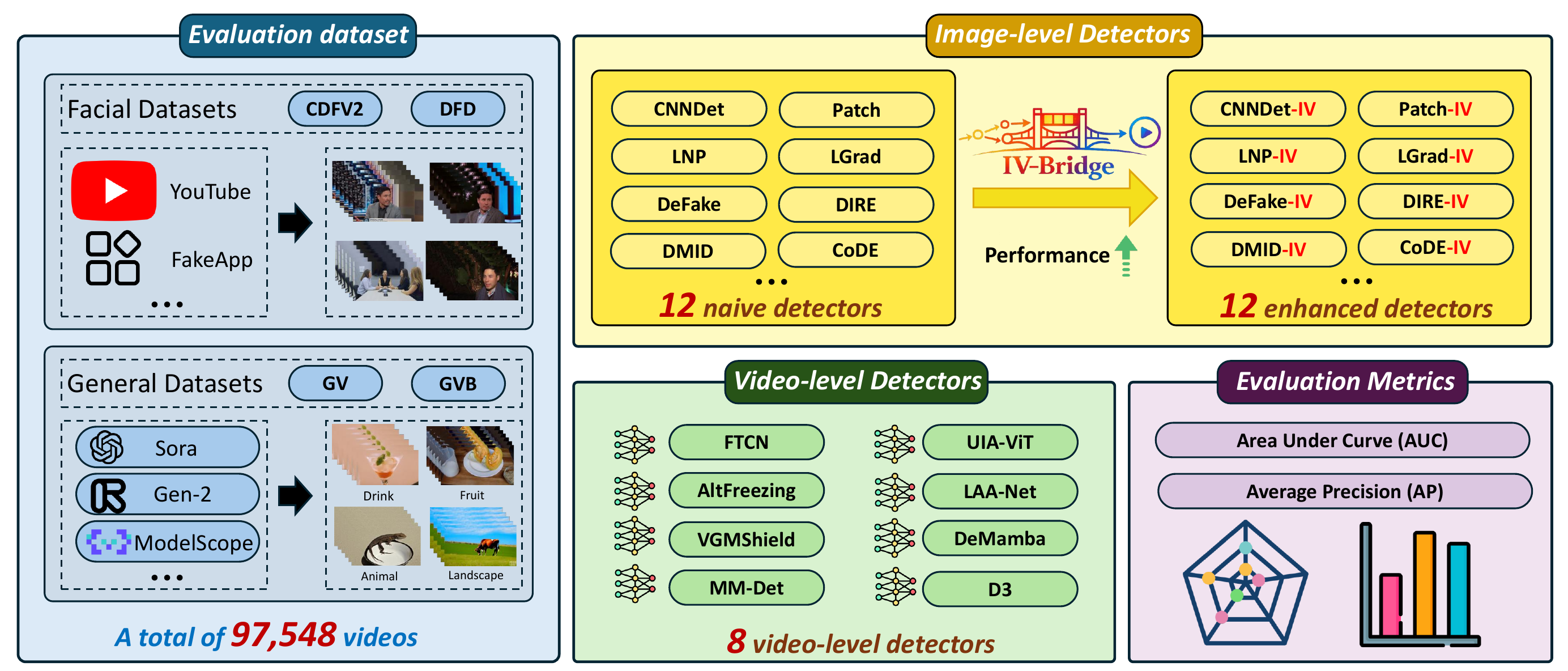}
\caption{An overview of \bench. \bench is designed to comprehensively evaluate the performance of deepfake video detectors, with particular emphasis on the scalability of image-level detectors.}
\label{fig:overview}
\end{figure*}

\begin{table*}[t]
\caption{Details of the evaluation datasets used in this paper.
We summarize the release years, dataset sizes, video sources, video content types, resolution, and frame rate (FPS).}
\label{tab:5_datasets}
\centering
\setlength{\tabcolsep}{6pt}
\begin{tabular}{cccccccc}
\toprule
\multirow{2}{*}{\textbf{Dataset}} & 
\multirow{2}{*}{\textbf{Year}} & 
\multirow{2}{*}{\makecell[c]{\textbf{Quantity} \\ \textbf{(Real/Fake)}}} 
& \multicolumn{2}{c}{\textbf{Video Source}} 
& \multirow{2}{*}{\makecell[c]{\textbf{Video} \\ \textbf{Content}}} 
& \multirow{2}{*}{\textbf{Resolution}} 
& \multirow{2}{*}{\textbf{FPS}} \\

\cmidrule(lr){4-5}

& & & \textbf{Real} & \textbf{Fake} & & & \\

\midrule

CDFV2~\cite{YXPHS20_1} & 2019 & 890/5,639
& YouTube 
& \makecell[c]{Improved basic DeepFake \\ makers (e.g., FakeApp)} 
& Facial & $256 \times 256$ & 30 \\
\midrule

DFD~\cite{dfd2019google} & 2019 & 363/3,068 
& Shot by actors 
& Unknown algorithms 
& Facial 
& \makecell[c]{$320 \times 320$--$3840 \times 2160$} 
& 15--24 \\
\midrule

GV~\cite{HYZZZYJHJWH24_1} & 2024 & 10,000/8,588 
& MSR-VTT~\cite{xu2016msr}
& \makecell[c]{Generated by diffusion and \\ autoregressive models} 
& General & $256 \times 256$ & 8--24 \\
\midrule

GVB~\cite{ZQMTYHXY25_1} & 2025 & 13,800/55,200 
& Vript~\cite{yang2024vript}
& \makecell[c]{Generated by diffusion and \\ autoregressive models} 
& General & $256 \times 256$ & 3--30 \\

\bottomrule
\end{tabular}
\end{table*}

\section{\bench}
We present \bench, a comprehensive benchmarking framework meticulously constructed to assess detector performance under complex deepfake video scenarios. 
It comprises four modules: Evaluation datasets, video-level detectors, image-level detectors, and evaluation metrics.
Next, we elaborate on each module in detail.

\subsection{Evaluation Datasets}
\bench collects two \textit{facial datasets}\footnote{Following~\cite{DNPPMAED24_1, YJDMF21_1}, we crop valid face regions from each frame.} and two \textit{general datasets} (including recent emerging AI-generated video datasets with diverse content).
The detailed statistics of these datasets are summarized in~\Cref{tab:5_datasets}.

First, the details of the facial datasets are as follows.
\begin{itemize}

\item \textbf{Celeb-DF v2 (CDFV2)}~\cite{YXPHS20_1} contains 5,639 high-quality synthetic videos and 890 real videos from YouTube.
The fake videos are generated using an improved DeepFake synthesis algorithm, which produces more realistic content and alleviates issues such as low-resolution faces, color mismatch, and temporal flickering.

\item \textbf{DeepFake Detection (DFD)}~\cite{dfd2019google} is a dataset released by Google \& JigSaw, comprising 3,068 fake videos and 363 real videos featuring professional actors.
Notably, the resolutions of the videos in DFD are not uniform, which is more conducive to evaluating the universality of the detectors.

\end{itemize}

Furthermore, we introduce the following two general datasets.

\begin{itemize}
    
\item \textbf{GenVideo (GV)}~\cite{HYZZZYJHJWH24_1} contains videos generated by 20 different video generation models (e.g., Gen2~\cite{runway2023gen2} and Sora~\cite{videoworldsimulators2024}), offering high visual realism and diversity.
We use its official test set containing 10,000 real videos from the MSR-VTT dataset~\cite{xu2016msr}, as well as 8,588 fake videos generated by diffusion VGMs (e.g., ModelScope~\cite{JHDYXS23_1}) or autoregressive VGMs (e.g., Videopoet~\cite{kondratyuk2023videopoet}), covering most mainstream generative paradigms and their advanced variants.

\item \textbf{GenVidBench (GVB)}~\cite{ZQMTYHXY25_1} is a challenging dataset comprising videos from eight SOTA text-to-video/image-to-video generators.
GVB ensures semantic richness and balanced distribution across three dimensions: actions, objects, and locations, making it a more challenging benchmark for detection tasks.
For evaluation, we use a directly accessible subset of 55,200 fake and 13,800 real videos from the full set.
The real videos are sampled from the Vript dataset~\cite{yang2024vript}, while the fake videos are also generated by both diffusion-based VGMs (e.g., SVD~\cite{ATSDMDYZVAVR23_1}, MuseV~\cite{MuseV}) and autoregressive-based VGMs (e.g., CogVideo~\cite{WMWXJ22_1}).
\end{itemize}

\begin{table*}[t]
\centering
\caption{Performance of deepfake detectors. 
The results of naive image-level detectors are the optimal aggregated results by one of the six modes.
The suffix ``-IV" indicates that the corresponding image-level detector has been enhanced using \ib.}
\label{tab:vft_all}
\setlength{\tabcolsep}{5.1pt}

\begin{tabular}{%
>{\centering\arraybackslash}p{0.8cm}%
>{\raggedright\arraybackslash}p{3.0cm}%
*{8}{>{\centering\arraybackslash}p{1.3cm}}%
}
\toprule
\multirow{3}{*}{\textbf{Type}} & \multirow{3}{*}{\textbf{Detector} }
& \multicolumn{4}{c}{\textbf{Facial Datasets}} & \multicolumn{4}{c}{\textbf{General Datasets}} \\
\cmidrule(lr){3-6} \cmidrule(lr){7-10}
& &  \multicolumn{2}{c}{\textbf{CDFV2}} 
& \multicolumn{2}{c}{\textbf{DFD}} 
& \multicolumn{2}{c}{\textbf{GVB}} 
& \multicolumn{2}{c}{\textbf{GV}} \\
\cmidrule(lr){3-4} \cmidrule(lr){5-6} \cmidrule(lr){7-8} \cmidrule(lr){9-10}
& & \textbf{AUC} $\uparrow$ & \textbf{AP} $\uparrow$& \textbf{AUC} $\uparrow$& \textbf{AP} $\uparrow$& \textbf{AUC} $\uparrow$& \textbf{AP} $\uparrow$& \textbf{AUC} $\uparrow$& \textbf{AP} $\uparrow$\\
\midrule
\multirow{9}{*}{\rotatebox[origin=c]{90}{Video-level}}
&FTCN~\cite{YJDMF21_1}   & 88.91\% & 98.61\% & \textbf{95.34}\%  & 99.30\% &71.32\%  & 80.75\%  &64.39\% & 57.82\% \\
&UIA-ViT~\cite{WQZQHCZN22_1}  & 87.17\% & 98.26\% & 84.45\% & 97.61\%  &76.52\% & 87.57\% &64.28\% & 58.49\% \\
&AltFreezing~\cite{ZJWWH23_1} & \textbf{92.06\%}  & 98.96\% & 77.54\% & 95.92\% &72.47\% & 81.23\% & 53.12  \%& 47.98 \%\\
&LAA-Net~\cite{DNPPMAED24_1}     & 85.40\% & 95.04\%  & 78.86\%  & 96.76\%  & 43.59\%  & 79.84\% & 59.39 \% & 62.95\% \\ 
\cmidrule{2-10}
&VGMShield~\cite{YYT24_1}  & 54.18\% & 90.92\%  &41.82 \% &83.05\%  & 82.50 \% & 94.95\% & 56.29\% & 61.34\%\\
&DeMamba~\cite{HYZZZYJHJWH24_1}  &54.35\% & 87.54\% & 51.79\% & 90.08\% & \textbf{88.69\%} &97.01\% &93.66\% &94.15 \%\\
&MM-Det~\cite{XXJQLXGX24_1}  & 50.69\% & 90.66\% & 51.49\% & 89.56\% & 63.49 \%  & 84.80\% & 43.62\% & 43.00\% \\
&D3~\cite{zheng2025d3}  & 49.00\% & 85.06\%  &52.70 \% & 90.44\% & 83.48\%  & 93.87\% & \textbf{94.37\%} & 94.16 \% \\
\midrule

\multirow{24}{*}{\rotatebox[origin=c]{90}{Image-level}}
& CNNDet~\cite{SORAA20_1}  & 52.62\% & 86.59 \% & 51.32\% & 88.54 \% & 61.59\%  & 86.55 \% & 72.55 \% & 70.20\%\\
& CNNDet-IV (\textbf{Ours}) & 85.62\% & 97.10 \% & 90.13 \% & 98.70\% & 82.77\% & 95.00\%  & 93.31\%  & 94.43\%\\
\cmidrule{2-10}
& LNP~\cite{BFXBWX22_1}  & 55.66\% & 87.03\% & 83.94\% & 97.24\% & 64.24\% & 85.39\% & 13.12\% & 30.56\% \\
& LNP-IV (\textbf{Ours}) & 70.45\%  & 93.16\% & 81.48\% & 97.34\% & 90.54\% & 96.72\% &94.96\% & 95.73\%\\
\cmidrule{2-10}
& Patch~\cite{LDSP20_1}  & 68.71\% & 91.60\% & 76.73\% & 96.26\% & 65.72\% & 88.24\% & 51.54\% & 54.47\% \\
& Patch-IV (\textbf{Ours}) & 87.92\% & 97.67\% & 91.40\% & 99.19\% & 85.51\%& 94.73\% & 92.21\% & 92.90\%\\
\cmidrule{2-10}
& LGrad~\cite{CYSGY23_1}  & 53.75\% & 86.98\% & 58.43\% & 90.68\% &58.54\% & 84.32\%& 28.45\% & 36.21\%\\
& LGrad-IV (\textbf{Ours}) & 73.26\% & 93.40\% & 83.37\% & 96.95\% & 72.21\% & 89.80\% & 90.20\% & 89.95\%\\
\cmidrule{2-10}
& DeFake~\cite{ZZNY23_1}  & 52.14\% & 87.24\% & 53.67\% & 90.54\% & 70.55\% & 90.06\% & 84.55\% & 83.45\% \\
& DeFake-IV (\textbf{Ours}) & 68.83\% & 92.52\% & 77.45\% & 96.27\% & 87.65\%& 96.60\%& 94.56\% & 94.61\%\\
\cmidrule{2-10}
& DIRE~\cite{ZJWWHHH23_1}  & 50.47\% & 85.93\% & 47.62\%  & 88.88\% & 60.57\% & 84.89\% & 66.02\% & 64.28\%\\
& DIRE-IV (\textbf{Ours}) & 73.55\% & 93.17\% & 78.04\% & 95.82\% &81.80\% & 93.80\% & 95.48\% & 94.76\%\\
\cmidrule{2-10}
& DMID~\cite{RDGGKL23_1}  & 50.94\% & 85.35\% & 57.01\% & 90.56\% & 92.02\% & 97.75\% & 99.37\% & 99.39\% \\
& DMID-IV (\textbf{Ours}) & 81.47\% & 95.94\% & 78.96\% & 96.67\% & 84.08\%& 94.93\% & 90.60\% & 89.79\%\\
\cmidrule{2-10}
& CoDE~\cite{LFMLAR24_1}  & 43.19\% & 83.13\% & 43.70\% & 87.26\% & 67.92\% & 89.05\% &73.43\% & 68.19\%\\
& CoDE-IV (\textbf{Ours}) & 79.30\% & 95.45\% & 82.40\% & 97.33\% & \textbf{99.56}\% & 98.85\% & \textbf{96.32}\% & 96.66\%\\
\cmidrule{2-10}
& DRCT~\cite{chen2024drct} & 54.04\% & 87.14\% & 56.30\% & 91.57\% & 90.01\% & 97.14\% & 97.65\% & 97.57\% \\
& DRCT-IV (\textbf{Ours}) & \textbf{90.81}\% & 98.33\% & 94.49\% &  99.27\% & 87.22\% & 96.10\% & 96.17\% & 97.76\%\\
\cmidrule{2-10}
& UniFD~\cite{UYJ23_1}  & 65.52\% & 91.76\% & 77.32\% & 96.72\% & 73.09\% & 91.38\% & 91.97\% & 91.59\%\\
& UniFD-IV (\textbf{Ours}) & 69.31\% & 92.81\% & 81.91\% & 97.17\%  & 76.51\% &92.61\% & 93.48\% & 93.05\%\\
\cmidrule{2-10}
& NPR~\cite{CYSGPY24_1}  & 55.46\% & 87.28\% & 66.85\% & 94.00\% & 48.25\% & 80.54\% & 60.00\%& 63.02\%\\
& NPR-IV (\textbf{Ours}) & 64.33\% & 91.65\% &78.83\% &96.82\% & 88.65\% & 95.68\% & 88.88\% & 90.21\% \\
\cmidrule{2-10}
& RINE~\cite{CS24_1}  & 63.87\% & 90.27\% & 78.53\% & 96.79\% & 82.03\% & 94.39\% & 96.21\% & 96.04\% \\
& RINE-IV (\textbf{Ours}) & 85.87\% & 96.76\% & \textbf{96.59}\% & 99.54\% & 98.99\%&99.68\% & 93.76\%& 94.54\%\\
\bottomrule
\end{tabular}
\end{table*}

\subsection{Video-level Detectors}
\bench considers eight representative video-level deepfake detectors, which contain four widely used \textit{face-focused} detectors (LAA-Net~\cite{DNPPMAED24_1}, FTCN~\cite{YJDMF21_1}, UIA-ViT~\cite{WQZQHCZN22_1}, and AltFreezing~\cite{ZJWWH23_1}) and four \textit{general-purpose} detectors (MM-Det~\cite{XXJQLXGX24_1}, VGMShield~\cite{YYT24_1}, DeMamba~\cite{HYZZZYJHJWH24_1}, and D3~\cite{zheng2025d3}).
These detectors are summarized in~\Cref{tab:video_level_deepfake_detection_methods}.
We employ officially released model weights for all detectors except DeMamba. 
Since the pretrained parameters for DeMamba are not publicly available, we train DeMamba following its official implementation details.

\subsection{Image-level Detectors}
\bench includes twelve advanced image-level deepfake detectors, which are listed in~\Cref{tab:13_image_deepfake_detectors}.
The selection is guided by three criteria.
First, we consider the training data domain, covering detectors trained on GAN-generated data, diffusion-generated data, or both.
Second, we include models with diverse backbone architectures such as ResNet, Vision Transformer, and so on.
Third, we account for different training paradigms:
Some methods explicitly model forgery artifacts (e.g., LNP~\cite{BFXBWX22_1}, LGrad~\cite{CYSGY23_1}), while others follow a data-driven paradigm without handcrafted priors (e.g., CNNDet~\cite{SORAA20_1}).
Similarly, we directly use the officially released model weights for all methods with the exception of LNP. 
Owing to the lack of publicly available pretrained weights for LNP, we generate its model weights in accordance with the official implementation.
Note that these detectors, which utilize the official implementations directly, are referred to as naive image-level detectors.
Additionally, in~\Cref{sec:ivbridge}, we will introduce an enhancement framework, named \ib, to uniformly augment these twelve naive detectors.

\subsection{Evaluation Metrics}
Following~\cite{kang2025hidf,DNPPMAED24_1,zheng2025d3}, \bench adopts Area Under the Receiver Operating Characteristic Curve (AUC) as the primary evaluation metric, and additionally reports Average Precision (AP). 
Both metrics are threshold-independent and provide a comprehensive assessment of detector performance.
AUC measures performance across different decision thresholds by characterizing the trade-off between the true positive rate and false positive rate, reflecting the overall ranking ability of a detector.
AP summarizes the Precision-Recall curve and emphasizes the detection quality of the positive class, focusing more on detection precision and recall.

\section{Benchmarking Video-level Detectors}
\label{sec:bench_video_level}
To answer \textbf{RQ1}, we evaluate eight state-of-the-art video-level detectors on the evaluation datasets of \bench. 
The results are reported in~\Cref{tab:vft_all}.

\mypara{Overall Performance} 
We begin by analyzing the overall performance across the four evaluation datasets.
FTCN achieves the best overall performance, with a mean AUC of 79.99\% and a mean AP of 84.12\%.
On facial datasets, FTCN performs particularly well on DFD, reaching an AUC of 95.34\% and an AP of 99.30\%, and also shows competitive results on CDFV2 (an 88.91\% AUC and a 98.61\% AP).
For general video datasets, DeMamba excels on GVB while D3 leads on GV.
Overall, face-specialized detectors dominate on facial content, whereas general-purpose detectors adapt better to non-face data. 
Nevertheless, even within their preferred domains, some detectors show suboptimal performance. 
For instance, AltFreezing achieves only 77.54\% AUC on DFD, and VGMShield and MM-Det drop to 56.29\% and 43.62\% AUC on GV, indicating substantial room for improvement in video-level deepfake detection.

\mypara{Generalization Performance} 
Cross-domain evaluation reveals that detectors' performance heavily depends on the target data domain. 
Facial detectors like FTCN and UIA-ViT maintain high AUC and AP on face datasets but drop significantly on general video content. 
Conversely, general-purpose detectors, such as DeMamba and D3, perform well on GV and GVB but exhibit limited performance on facial datasets, with AUC dropping from around 94\% to approximately 50\%.
These results indicate that current video-level detectors lack sufficient cross-domain generalization, highlighting the need for future designs that are more effective and generalizable across diverse video content.

\begin{tcolorbox}[colback=black!3!white,enhanced jigsaw,breakable]
\textbf{Answer to RQ1:}
Despite achieving excellent performance (e.g., >90\% AUC) on specific datasets, current mainstream detectors suffer from poor generalization. 
For instance, face-specialized detectors exhibit a significant performance drop on challenging general datasets. 
Similarly, detectors that excel on general datasets only achieve around 50\% AUC on facial benchmarks. 
In summary, current mainstream video-level detectors fall short of delivering universally high performance.
\end{tcolorbox}

\section{Benchmarking Naive Image-level Detectors}
\label{sec:benchmark_naive}
We investigate the performance of naive image-level detectors in the video domain from two perspectives to answer \textbf{RQ2}.
First, we assess their performance on individual video frames. 
Second, we evaluate their scalability by employing different strategies to aggregate frame-level detection results.

\subsection{Performance on Video Frames}
\mypara{Dataset Construction}
To thoroughly analyze the image-level detectors' performance on video frames and their divergence from the original image domain, we additionally construct a carefully curated dataset named FakeGenImage.
FakeGenImage contains 52,160 fake images generated by 20 distinct image synthesis models, covering diverse generation paradigms including GAN-based models~\cite{karras2017celebhq,stylegan,stylegan2,biggan,cyclegan,choi2018stargan,gaugan}, diffusion-based models~\cite{dhariwal2021guided,Rombach_2022_CVPR,nichol2021glide,ramesh2021dalle}, and other representative generation approaches~\cite{FF++,seeingindark,dai2019san,chen2017crn,li2019imle}. 
It also includes 52,169 real images collected from commonly used image and face datasets, including LSUN~\cite{yu2015lsun}, ImageNet~\cite{russakovsky2015imagenet}, CelebA~\cite{liu2015celeba}, CelebA-HQ~\cite{karras2017celebhq}, COCO~\cite{lin2014microsoftcoco}, FaceForensics++~\cite{FF++}, and LAION~\cite{schuhmann2021laion}.
These images have been widely adopted in prior studies~\cite{UYJ23_1,CYSGPY24_1,CS24_1} as standard benchmarks for evaluating the performance of image-level deepfake detectors.

\mypara{Evaluation Results}
The performance of the naive image-level detectors on FakeGenImage and video frames from \bench is shown in \Cref{tab:frame_level_result_auc_ap_main}. 
Based on the AUC results, we observe that the performance of 10 out of 12 detectors declines on video frames, with severe degradation in some cases (e.g., NPR drops from 92.17\% to 55.38\%). 
This degradation is largely due to differences between fake images and video frames, such as motion blur, resolution variations, and other visual distortions introduced during video generation and encoding. 
Furthermore, while Patch and Defake show a slight performance improvement on frames, their AUCs are still below 60\%, and both struggle on the original image domain. 
The best-performing detector on video frames is UniFD, yet it only achieves an AUC of 71.78\%. 
Overall, these results demonstrate the severe inadequacy of naive image-level detectors for video frame detection.

\begin{table}[t]
\centering
\caption{The detection results of naive image-level detectors on deepfake images and deepfake video frames. }
\label{tab:frame_level_result_auc_ap_main}
\setlength{\tabcolsep}{9pt}
\begin{tabular}{l c c c c}
\toprule
\multirow{2}{*}{\textbf{Detector}} 
& \multicolumn{2}{c}{\textbf{FakeGenImage}} 
& \multicolumn{2}{c}{\textbf{\bench}} \\
\cmidrule(lr){2-3} \cmidrule(lr){4-5}
& \textbf{AUC} $\uparrow$ & \textbf{AP} $\uparrow$ & \textbf{AUC} $\uparrow$ & \textbf{AP} $\uparrow$ \\
\midrule
CNNDet~\cite{SORAA20_1} & 83.28\% & 82.54\% & 48.37\% & 64.84\%  \\
LNP~\cite{BFXBWX22_1} & 71.75\% & 69.40\% & 51.02\% & 66.78\%  \\
Patch~\cite{LDSP20_1} &  46.16\% & 52.52\% & 59.62\%  & 68.82\% \\
LGrad~\cite{CYSGY23_1} & 85.45\% & 85.23\% & 45.81\% & 64.47\% \\
DeFake~\cite{ZZNY23_1} & 46.76\% & 49.96\% & 59.66\% & 73.32\% \\
DIRE~\cite{ZJWWHHH23_1} & 69.52\% & 67.83\% & 47.71\% & 64.11\% \\
DMID~\cite{RDGGKL23_1} & 95.50\% & 94.86\% & 67.44\% & 86.69\% \\
CoDE~\cite{LFMLAR24_1} & 67.49\% & 64.26\% & 50.00\% & 66.70\% \\
DRCT~\cite{chen2024drct} & 77.32\% & 76.64\% & 71.56\% & 88.36\% \\
UniFD~\cite{UYJ23_1} & 93.73\% & 93.67\%  & 71.78\% & 79.98\%\\
NPR~\cite{CYSGPY24_1} & 92.17\% & 90.95\% & 55.38\% & 69.55\% \\
RINE~\cite{CS24_1} & 98.80\% & 98.80\% & 71.11\% & 83.01\% \\
\bottomrule
\end{tabular}
\end{table}

\begin{figure*}[t]
\centering
\includegraphics[width=1.0\textwidth]{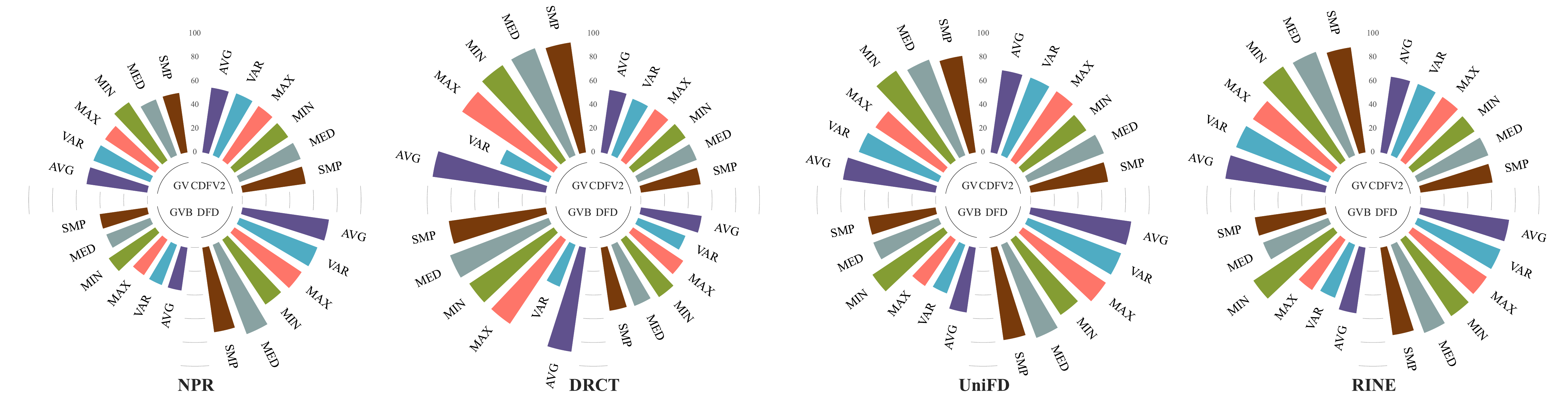}
\caption{The video-level performances of naive image-level detectors using six aggregation modes.}
\label{fig:4_of_12_ID_modes_wo_vft}
\end{figure*}

\subsection{Performance on Videos}
\label{sec:frame_to_video}
For these naive image-level detectors, we further explore several lightweight strategies to aggregate frame-level predictions into a video-level decision.

\mypara{Aggregation Strategies}
We consider six distinct frame-to-video aggregation strategies: Sampling Mode (SMP), Averaging Mode (AVG), Maximum Mode (MAX), Minimum Mode (MIN), Median Mode (MED), and Variance Mode (VAR). 
For instance, the SMP mode randomly selects a single frame from the video and uses its score as the video-level prediction. Detailed definitions of all modes are provided in Appendix~\ref{app:f2v_strategy}.
\Cref{fig:4_of_12_ID_modes_wo_vft,fig:8_of_12_video-level-mode-wo_vft} depict the detection performance of different strategies for various detectors. It reveals that the optimal aggregation strategy differs across detectors. 
Furthermore, for some detectors, performance varies significantly between different strategies.
For example, RINE achieves an AUC of only 68.60\% under the VAR mode, whereas its performance improves to 80.16\% when the MIN strategy is adopted.

\mypara{Video-level Performance}
For ease of comparison, the video-level detection results after aggregation are also reported in~\Cref{tab:vft_all}. 
The reported results are the optimal aggregated results by default.
The results demonstrate that aggregating frame-level predictions into video-level scores significantly improves detection performance. 
For instance, RINE achieves an AUC of 71.11\% at the frame level, which increases to 80.16\% after video-level aggregation, representing an improvement of approximately 10\% and surpassing even the best video-level detector, FTCN (79.99\% AUC).
These findings further indicate that previous benchmarks~\cite{deepfakebench}, which directly compare frame-level results of image-level detectors with video-level detectors, are neither reasonable nor fair.
Moreover, they highlight that leveraging image-level detectors for deepfake video detection is an underexplored direction that warrants further investigation.

\begin{tcolorbox}[colback=black!3!white,enhanced jigsaw,breakable]
\textbf{Answer to RQ2:}  
The detection performance of naive image-level detectors significantly declines when applied to video frames compared to the image domain. However, through various aggregation methods, improved video-level detection results can be achieved, even surpassing current state-of-the-art video-level detectors. 
This fully demonstrates the potential of image-level detectors.
\end{tcolorbox}

\section{Improving the Applicability of Naive Image-level Detectors}
\label{sec:ivbridge}

Based on the findings of~\Cref{sec:benchmark_naive}, we introduce \ib, a systematic framework designed to strengthen the performance of image-level detectors in video scenarios.

\begin{figure*}[t]
\centering
\includegraphics[width=1\textwidth]{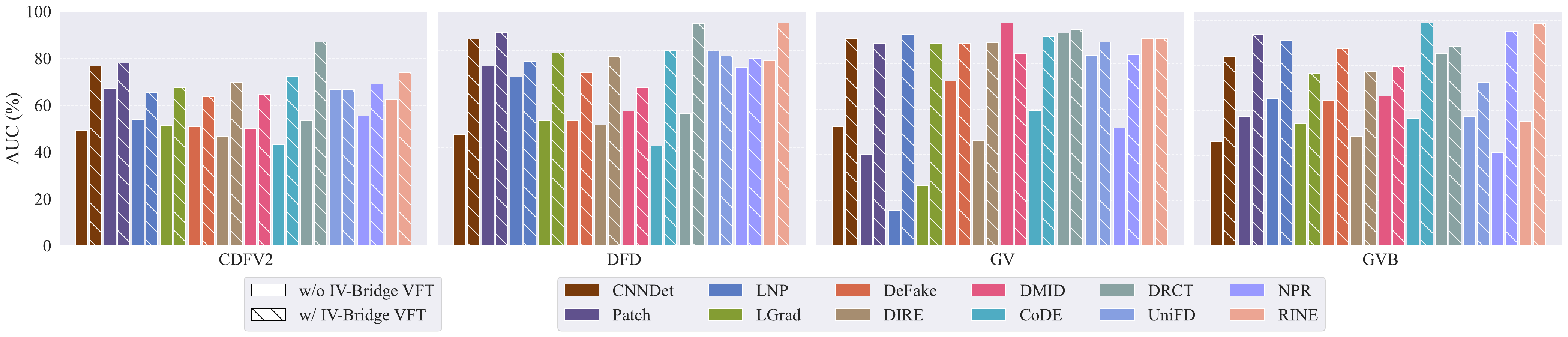}
\caption{Frame-level detection performance (AUC \%) of image-level detectors on \bench. 
“w/o VFT” denotes the original detectors, and “w/ VFT” stands for the enhanced detectors via VFT.}
\label{fig:frame-level-compare}
\end{figure*}

\subsection{Methodology of \ib}

\ib consists of a two-stage pipeline:
Video-Frame Fine-Tuning (VFT) and Multi-Mode Aggregation (MMA).

\mypara{Video-Frame Fine-Tuning (VFT)}
Since image-level detectors are only trained on forged images, they cannot fully capture the features of forged video frames due to distribution shifts.
To bridge the gap between images and video frames, \ib fine-tunes the naive image-level detector $\mathcal{M}_{img}$ with a fine-tuning dataset $\mathcal{D}_{ft}$ under a full-parameter setting to generate $\mathcal{M}_{img}^{VFT}$.

Given an image $I$, the outputs of $\mathcal{M}_{img}$ are primarily categorized into the following two categories.
Accordingly, we design distinct fine-tuning loss functions for each category.
\begin{itemize}
\item \textbf{Two-logit output.}  
The first category (e.g., DeFake, DRCT) outputs two scores indicating whether an image is real or fake, i.e.,
$\mathcal{M}_{img}: I \mapsto \mathbf{p}=(p^\text{real}, p^\text{fake})$.
We use the standard cross-entropy loss for fine-tuning:
\begin{equation*}
\mathcal{L}_{CE} = -\frac{1}{N}\sum_{i=1}^{N} \log p_i^{y_i},
\end{equation*}
where $y_i\in \{\text{real}, \text{fake}\}$ is the ground-truth label of the $i$-th frame and $p_i^{y_i}$ denotes the predicted probability for the ground-truth label $y_i$, and $N$ is the dataset size of $\mathcal{D}_{ft}$.

\item \textbf{Single-logit output.}  
The second category (e.g., CNNDet, NPR) outputs only the confidence score that the image is fake, i.e.,
$\mathcal{M}_{img}: I \mapsto p^\text{fake}$.
We use the following binary cross-entropy (BCE) function to fine-tune detectors:
\begin{equation*}
\mathcal{L}_{BCE} = -\frac{1}{N}\sum_{i=1}^{N}\big[y_i\log p_i^{\text{fake}} + (1-y_i)\log(1-p_i^{\text{fake}})\big],
\end{equation*}
where $y_i \in \{0(\text{real}), 1(\text{fake})\}$ is the ground-truth label.
\end{itemize}

\mypara{Multi-Mode Aggregation (MMA)} 
\ib further introduces a multi-mode aggregation strategy that adaptively fuses information from multiple frame-to-video modes in a learning-based manner.
Specifically, given $\mathcal{M}_{img}^{VFT}$ and a video with $T$ frames $\{f_i\}^T_{i=1}$, the detector outputs a forgery probability for each frame $f_i$:
\begin{equation*}
    p^{\text{fake}}_i = \mathcal{M}^{VFT}_{img}(f_i) \in [0, 1],~i=1, \dots, T.
\end{equation*}
Next, these frame-level scores are aggregated under six different frame-to-video aggregation mode functions $\phi_m$ as detailed in Appendix~\ref{app:f2v_strategy}, where $m \in \{\text{SMP}, \text{AVG}, \text{MAX}, \text{MIN}, \text{MED}, \text{VAR}\}$.
Each function maps the frame-level scores $\{p_i^{\text{fake}}\}_{i=1}^{T}$ to a video-level score:
\begin{equation*}
    P_m = \phi_m(p^{\text{fake}}_1, \cdots, p^{\text{fake}}_T).
\end{equation*}
Stacking the outputs of all modes yields the multi-mode video feature vector:
\begin{equation*}
\mathbf{P}_{V} = [P_{\text{SMP}}, P_{\text{AVG}}, P_{\text{MAX}}, P_{\text{MIN}}, P_{\text{MED}}, P_{\text{VAR}}]^\top.
\end{equation*}
Inspired by the use of lightweight classifiers in recent video detection studies~\cite{interno2026ai}, we employ a lightweight random forest classifier~\cite{DBLP:journals/ml/Breiman01} $\mathcal{C}_{\rm RF}(\cdot)$ trained on a dataset $\mathcal{D}_{vid}$ to map the resulting multi-mode features to a final video-level forgery probability:
\[
z_V = \mathcal{C}_{\rm RF}(\mathbf{P}_V), ~z_V \in [0,1].
\]
Through this learning-based aggregation, the model automatically captures the relative importance and interactions among different modes, allowing more flexible and effective video-level predictions compared to fixed single-mode aggregation.

\subsection{Experiments}

\mypara{Experimental Setups}
In \ib, we use a total of 197,533 real and 207,476 fake video frames, including 92,157 real and 92,146 fake frames from the FF++~\cite{FF++} (c23 version), a widely used face forgery dataset, and 105,376 real and 115,300 fake frames from the training split of GenVideo~\cite{HYZZZYJHJWH24_1}.
These data do not have any overlapping images with our \bench evaluation dataset.
For all image-level detectors, we generally follow their original hyperparameters during training.
For the MMA stage, the Random Forest classifier is configured with up to 800 trees and a maximum depth of 8.

\begin{figure}[t]
\centering
\begin{subfigure}{0.95\linewidth}
\centering
\includegraphics[width=0.95\linewidth]{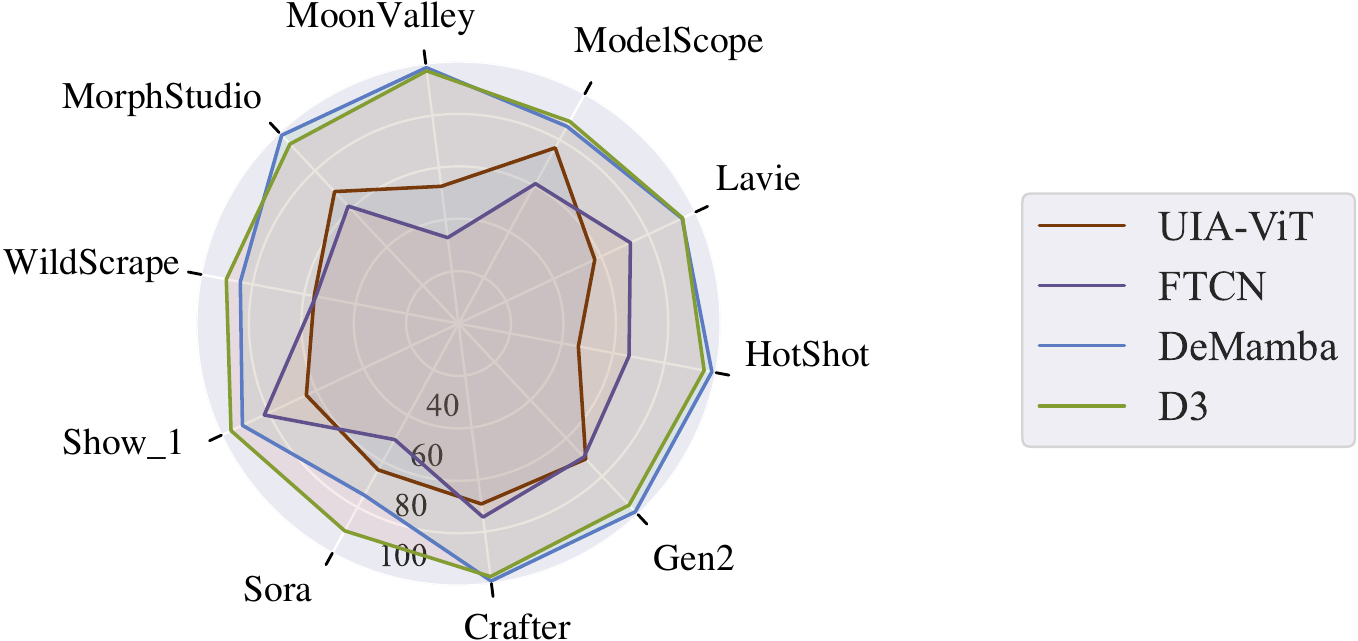}
\caption{Video-level Detectors.}
\label{fig:genvideo_10_datasets_radar_sub1}
\end{subfigure}
\begin{subfigure}{0.95\linewidth}
\centering
\includegraphics[width=0.95\linewidth]{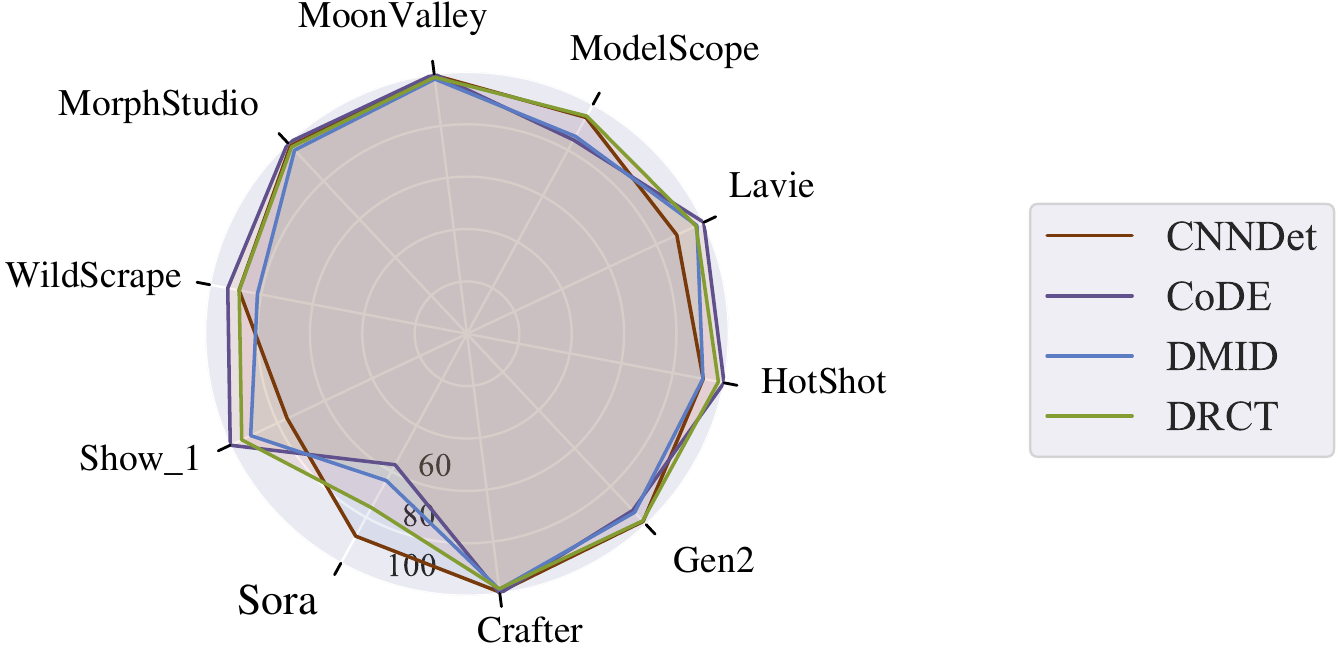}
\caption{IV-Bridge Enhanced Detectors.}
\label{fig:genvideo_10_datasets_radar_sub2}
\end{subfigure}
\caption{Results of cross-model detection performance (AUC \%) on GV dataset (Part-I).}
\label{fig:genvideo_10_datasets_radar}
\end{figure}

\mypara{Overall Performance}
Using \ib, we obtain twelve \ib-enhanced detectors.
\Cref{tab:vft_all} details the performance of each individual detector across the evaluation datasets of \bench, where the suffix ``-IV" indicates that the corresponding detector has been enhanced using \ib.
Compared with directly applying the original naive image-level detectors, the \ib-enhanced versions achieve substantial performance gains on all datasets.
For example, after \ib adaptation, DRCT-IV improves its AUC by around 18\% over the original DRCT.
\ib particularly boosts fully data-driven detectors like DRCT and RINE, which outperform artifact-specific ones, achieving an average AUC of 87.35\% versus 81.63\%.
More importantly, the enhanced image-level detectors significantly outperform the best video-level detector.
In terms of the performance across the full \bench dataset, the SOTA video-level detector FTCN attains the highest AUC of 79.99\% and AP of 84.12\%. 
However, among the twelve detectors enhanced by \ib, eleven detectors achieve performance exceeding that of FTCN, with RINE-IV reaching the highest performance of 93.80\% AUC and 97.63\% AP.
Moreover, \ib boosts performance on both facial datasets and general datasets. 
For instance, RINE-IV reaches 96.59\% AUC on DFD and CoDE-IV achieves 96.32\% AUC on GV, both surpassing the original SOTA video-level detector.
Notably, the enhanced CoDE-IV attains an AUC of 99.56\% on the GVB dataset, approaching near-perfect detection performance.

\mypara{Performance across Different VGMs}
Given the widespread accessibility of diverse video generation models (VGMs), we further analyze how detection performance varies across different generators.
The GV and GVB datasets contain videos generated by ten and four distinct models, respectively.
For the GV dataset, \Cref{fig:genvideo_10_datasets_radar} presents the performance of four representative detectors, while the results of the remaining detectors are shown in~\Cref{fig:genvideo_9_3_datasets_radar}.
Overall, the \ib-enhanced image-level detectors demonstrate strong generalization across videos produced by diverse VGMs.
Nevertheless, noticeable performance variations still exist among different generation models.
Specifically, videos generated by Sora are more challenging to detect, yielding a mean AUC of 72.76\%, whereas those generated by Crafter are the easiest, achieving an AUC of 98.32\%.
For the GVB dataset, as shown in~\Cref{fig:gvb_13_7_datasets_radar}, videos generated by SVD and MuseV are also relatively more difficult.
This can be attributed to the fact that SVD and MuseV generate videos by taking a real image as the initial frame, resulting in more realistic visual content, while videos officially released by Sora exhibit notably high visual quality.
Compared with dedicated video-level detectors, \ib-enhanced image-level detectors achieve a higher mean AUC of 95.51\% across all 14 generation models, surpassing the SOTA video-level detector, which attains an AUC of only 91.01\%.

\mypara{Effectiveness of VFT}
To evaluate the effectiveness of Video-Frame Fine-Tuning (VFT), we compare the frame-level detection performance of image-level detectors on the four \bench subsets (CDFV2, DFD, GV, and GVB) before and after applying VFT. The results are illustrated in~\Cref{fig:frame-level-compare}.
As shown in the figure, all twelve image-level detectors exhibit substantial improvements in frame-level AUC on \bench videos after VFT. 
This indicates that detectors trained solely on manipulated still images struggle to capture the visual characteristics of forged video frames. 
Fine-tuning on video frames effectively mitigates the distribution gap between images and video frames, enabling detectors to better adapt to the video domain.
These results further confirm that forged images and forged video frames exhibit different forgery patterns and visual statistics, and highlight the necessity of video-specific adaptation for improving the applicability of image-level detectors in deepfake video detection tasks.

\mypara{Deployment Cost}
\Cref{tab:efficiency_comparison_results} summarizes the model parameters (M), inference time (ms), and detection performance (\%) of detectors.
All experiments are conducted on a single NVIDIA A800 GPU with 80GB VRAM.
First of all, we can observe that three of the twelve \ib-enhanced detectors (NPR, Patch, and CoDE) outperform the SOTA video-level detector FTCN across all three key metrics: detection performance, model size, and inference time.
Notably, Patch-IV, with a parameter count of merely 4.34M, is more lightweight than all video-level detectors listed. 
However, it achieves superior performance with an AUC of 89.26\% and an AP of 96.12\%, which is notably higher than all listed video-level detectors. 
In terms of inference speed, CNNDet-IV ranks first among image-level models at 4.24ms, delivering an AUC of 87.96. 
In contrast, the fastest video-level detector, UIA-ViT, requires 7.59ms but attains a significantly lower AUC of 78.11\%.

\begin{table}[t]
\centering
\caption{Deployment costs of different detectors.
}
\label{tab:efficiency_comparison_results}
\setlength{\tabcolsep}{1.5pt}
\begin{tabular}{c l r r c c}
\toprule
\multirow{2}{*}{\textbf{Type}} & 
\multirow{2}{*}{\textbf{Detector}} & 
\multirow{2}{*}{\makecell[c]{\textbf{\#Param} \\ \textbf{(M)}}} & 
\multirow{2}{*}{\makecell[c]{\textbf{Time} \\ \textbf{(ms)}}} & 
\multicolumn{2}{c}{\textbf{Performance}} \\
\cmidrule(lr){5-6}
& & & & \textbf{AUC(\%)}$\uparrow$ & \textbf{AP(\%)}$\uparrow$ \\
\midrule

\multirow{8}{*}{\rotatebox[origin=c]{90}{Video-level}}

& FTCN~\cite{YJDMF21_1} & \parabarblue{14}{14.77} & \timebargreen{11}{11.97} & 79.99 & 84.12 \\
& UIA-ViT~\cite{WQZQHCZN22_1} & \parabarblue{85}{85.80} & \timebargreen{7}{\textbf{7.59}} & 78.11 & 85.48 \\
& AltFreezing~\cite{ZJWWH23_1} & \parabarblue{27}{27.23} & \timebargreen{33}{33.44} & 73.80 & 81.02 \\
& LAA-Net~\cite{DNPPMAED24_1} & \parabarblue{9}{\textbf{9.74}} & \timebargreen{22}{22.43} & 66.81 & 83.65 \\
& VGMShield~\cite{YYT24_1} & \parabarblue{64}{64.98} & \timebargreen{14}{14.92} & 58.70 & 82.57 \\
& DeMamba~\cite{HYZZZYJHJWH24_1} & \parabarblue{127}{127.27} & \timebargreen{344}{344.63} & 72.12 & 92.20 \\
& MM-Det~\cite{XXJQLXGX24_1} & \parabarblue{203}{203.62} & \timebargreen{76}{76.9} & 52.32 & 77.01 \\
& D3~\cite{zheng2025d3} & \parabarblue{121}{121.25} & \timebargreen{24}{24.68} & 69.89 & 90.88 \\
\midrule

\multirow{12}{*}{\rotatebox[origin=c]{90}{Image-level-enhanced}}

& CNNDet-IV & \parabarblue{23}{23.51} & \timebargreen{4}{\textbf{4.24}} & 87.96 & 96.31 \\
& LNP-IV & \parabarblue{26}{26.35} & \timebargreen{128}{128.07} & 84.36 & 95.74 \\
& Patch-IV & \parabarblue{4}{4.34} & \timebargreen{9}{9.69} & 89.26 & 96.12 \\
& LGrad-IV & \parabarblue{46}{46.56} & \timebargreen{138}{138.62} & 79.76 & 92.53 \\
& DeFake-IV & \parabarblue{375}{375.61} & \timebargreen{1772}{1772.7} & 82.12 & 95.00 \\
& DIRE-IV & \parabarblue{550}{552.67} & \timebargreen{16000}{16481.48} & 82.22 & 94.39 \\
& DMID-IV & \parabarblue{23}{23.51} & \timebargreen{84}{84.81} & 83.78 & 94.33 \\
& CoDE-IV & \parabarblue{5}{5.52} & \timebargreen{4}{4.32} & 89.40 & 97.07 \\
& DRCT-IV & \parabarblue{88}{88.62} & \timebargreen{26}{26.08} & 92.17 & 97.87 \\
& UniFD-IV & \parabarblue{427}{427.62} & \timebargreen{116}{116.18} & 80.30 & 93.91 \\
& NPR-IV & \parabarblue{1}{\textbf{1.44}} & \timebargreen{6}{6.87} & 80.17 & 93.59 \\
& RINE-IV & \parabarblue{433}{433.94} & \timebargreen{18}{18.37} & 93.80 & 97.63 \\

\bottomrule
\end{tabular}
\end{table}

\mypara{Performance on Short-term Forgery}
Sparse manipulations, where only a small portion of video frames are forged, pose a more challenging detection setting.
To examine this case, we construct a short-term forgery test set based on the facial dataset CDFV2 by keeping only 20\% forged frames in each fake video.
We evaluate this setting on face-specialized video-level detectors and all IV-Bridge-enhanced image-level detectors.
As shown in Table~\ref{tab:short-term}, all detectors suffer clear performance drops compared with the original CDFV2 setting, confirming that sparse temporal forgery remains challenging for current detectors. 
Nevertheless, several IV-Bridge-enhanced image-level detectors remain competitive.
Patch-IV achieves the best AUC of 77.15\% with a relatively moderate AUC drop of 10.77 percentage points, while DRCT-IV obtains 74.09\% AUC.
These results suggest that IV-Bridge can still maintain superior detection performance under sparse forged-frame scenarios.

\begin{table}[t]
\centering
\caption{Performance under the short-term forgery setting on CDFV2. Only 20\% frames in each fake video are forged. $\Delta$AUC and $\Delta$AP denote the absolute performance drops compared with the original CDFV2 setting.}
\label{tab:short-term}
\setlength{\tabcolsep}{4.5pt}
\begin{tabular}{c l c c c c}
\toprule
\textbf{Type} & \textbf{Detector} &  
\textbf{AUC}$\uparrow$ & 
\textbf{AP}$\uparrow$ & 
\textbf{$\Delta$AUC}$\downarrow$ & 
\textbf{$\Delta$AP}$\downarrow$ \\
\midrule

\multirow{4}{*}{\rotatebox[origin=c]{90}{Video-level}}

& FTCN~\cite{YJDMF21_1} & 51.17\% & 85.59\% & 37.74\% & 13.02\% \\
& UIA-ViT~\cite{WQZQHCZN22_1} & 58.69\% & 88.69\% & 28.48\% & 9.57\% \\
& AltFreezing~\cite{ZJWWH23_1} & 56.27\% & 88.17\% & 35.79\% & 10.79\% \\
& LAA-Net~\cite{DNPPMAED24_1} & \textbf{67.15\%} & 90.50\% & 18.25\% & 4.54\% \\
\midrule

\multirow{12}{*}{\rotatebox[origin=c]{90}{Image-level-enhanced}}

& CNNDet-IV & 64.93\% & 91.51\% & 20.69\% & 5.59\% \\
& LNP-IV & 56.64\% & 88.68\% & 13.81\% & 4.48\% \\
& Patch-IV & \textbf{77.15\%} & 94.82\% & 10.77\% & 2.85\% \\
& LGrad-IV & 60.33\% & 89.21\% & 12.93\% & 4.19\% \\
& DeFake-IV & 50.73\% & 86.01\% & 18.10\% & 6.51\% \\
& DIRE-IV & 57.87\% & 89.15\% & 15.68\% & 4.02\% \\
& DMID-IV & 55.55\% & 88.52\% & 25.92\% & 7.42\% \\
& CoDE-IV & 43.77\% & 84.55\% & 35.53\% & 10.90\% \\
& DRCT-IV & 74.09\% & 94.58\% & 16.72\% & 3.75\% \\
& UniFD-IV & 61.17\% & 91.06\% & 8.14\% & 1.75\% \\
& NPR-IV & 41.03\% & 81.40\% & 23.30\% & 10.25\% \\
& RINE-IV & 46.08\% & 84.81\% & 39.79\% & 11.95\% \\

\bottomrule
\end{tabular}
\end{table}

\begin{tcolorbox}[colback=black!3!white,enhanced jigsaw,breakable]
\textbf{Answer to RQ3:}  
The proposed \ib boosts the performance of naive image-level detectors in deepfake video detection, with 11 of the 12 enhanced detectors exceeding the state-of-the-art video-level detector (e.g., FTCN). 
This demonstrates the stronger generalization, computational lightweightness, and practical effectiveness of \ib in adapting image-level detectors to video-level tasks.
\end{tcolorbox}

\section{Conclusion}
In this work, we introduce \bench, a benchmark dedicated to systematically evaluating the applicability of image-level deepfake detectors for video-level deepfake detection. 
\bench contains $97,548$ videos and supports a comprehensive evaluation of eight video-level detectors and twelve representative image-level detectors. 
Furthermore, we present \ib, a two-stage framework consisting of Video-Frame Fine-Tuning (VFT) and Multi-Mode Aggregation (MMA), designed to enhance image-level detectors for video tasks. 
Our evaluation reveals that: (1) current video-level detectors still leave substantial room for performance improvement; 
(2) naive image-level detectors suffer significant performance drops when applied to video frames, while appropriate frame aggregation can partially mitigate this gap; 
and (3) image-level detectors can be significantly improved with \ib, surpassing state-of-the-art video-level methods with lower computational cost and stronger generalization across different video generation models.
Overall, \bench establishes a rigorous benchmark for deepfake video detection and, together with \ib, provides a practical and efficient solution for adapting image-level detectors to video scenarios, offering valuable guidance for future research in deepfake video detection.

\section*{Acknowledgement}
We sincerely thank the reviewers for their valuable feedback. 
This work is supported by the National Natural Science Foundation of China (No.62402273) and the Fundamental and Interdisciplinary Disciplines Breakthrough Plan of the Ministry of Education of China under Grant JYB2025XDXM114.
Tianshuo Cong is also with the Shandong Key Laboratory of  Artificial Intelligence Security.

\bibliographystyle{ACM-Reference-Format}
\bibliography{refs}

\appendix

\section{Aggregation Strategies}
\label{app:f2v_strategy}
Given an image-level detector $\mathcal{M}_{img}$ and a video $V = \{f_1, \dots, f_N\}$, the detector outputs a fake confidence score $p_i^{\text{fake}}$ for each frame $f_i$.
We evaluate six frame-to-video integration mode functions $\phi_m$, where $ m \in \{\text{SMP, AVG, MAX, MIN, MED, and VAR}\}$ to obtain video-level prediction $P_m$.
The detailed definitions of these six modes are provided as follows.
\begin{itemize}
    \item \textbf{Sampling Mode (SMP).}  
    The SMP mode randomly selects a single frame $F_i$ from video $V$ and uses its score as the video-level prediction:
    \begin{equation*}
    P_{\text{SMP}} = \phi_{\text{SMP}}(\mathcal{M}_{img}, V) = p_i^{\text{fake}}, 
    \end{equation*}
    
    \item \textbf{Averaging Mode (AVG).}  
    The AVG mode computes the mean score across all frames:
    \begin{equation*}
    P_{\text{AVG}} = \phi_{\text{AVG}}(\mathcal{M}_{img}, V) = \frac{1}{N}\sum^N_{i=1}p_i^{\text{fake}}. 
    \end{equation*}

    \item \textbf{Maximum Mode (MAX).}  
    The MAX mode uses the largest frame-level score as the final video-level prediction:
    \begin{equation*}
    P_{\text{MAX}} = \phi_{\text{MAX}}(\mathcal{M}_{img}, V) = \max_{i = 1, \dots, N} p_i^{\text{fake}}. 
    \end{equation*}

    \item \textbf{Minimum Mode (MIN).}  
    The MIN mode adopts the smallest frame score:
    \begin{equation*}
    P_\text{MIN} = \phi_{\text{MIN}}(\mathcal{M}_{img}, V) = \min_{i = 1, \dots, N} p_i^{\text{fake}}. 
    \end{equation*}

    \item \textbf{Median Mode (MED).}  
    The MED mode first sorts the frame-level scores in ascending order: 
    $$\tilde{p}^{\text{fake}}_{(1)} \leq \cdots \leq \tilde{p}^{\text{fake}}_{(N)}, $$
    where $\tilde{p}^{\text{fake}}_{(k)}$ represents the $k$-th order score.
    The median value is then computed as the final video-level score.   

    \[
    P_{\text{MED}} = \phi_{\text{MED}}(\mathcal{M}_{img}, V)=
    \begin{cases}
    \tilde{p}^{\text{fake}}_{\left(\frac{N+1}{2}\right)}, & \text{if $N$ is odd},\\[2mm]
    \dfrac{\tilde{p}^{\text{fake}}_{\left(\frac{N}{2}\right)} + \tilde{p}^{\text{fake}}_{\left(\frac{N}{2}+1\right)}}{2}, & \text{if $N$ is even}.
    \end{cases}
    \]

    \item \textbf{Variance Mode (VAR).}  
    The VAR mode computes the variance of frame-level scores:
    \begin{equation*}
    P_{\text{VAR}} = \phi_{\text{VAR}}(\mathcal{M}_{img}, V) = \frac{1}{N}\sum^N_{i = 1}(p_i^{\text{fake}} - \bar{p}^{\text{fake}})^2,~\bar{p}^{\text{fake}} = \frac{1}{N}\sum^N_{i=1}p_i^{\text{fake}}. 
    \end{equation*}     

\end{itemize}

\begin{figure}[t]
    \centering

    \begin{subfigure}{0.9\linewidth}
        \centering
        \includegraphics[width=0.9\linewidth]{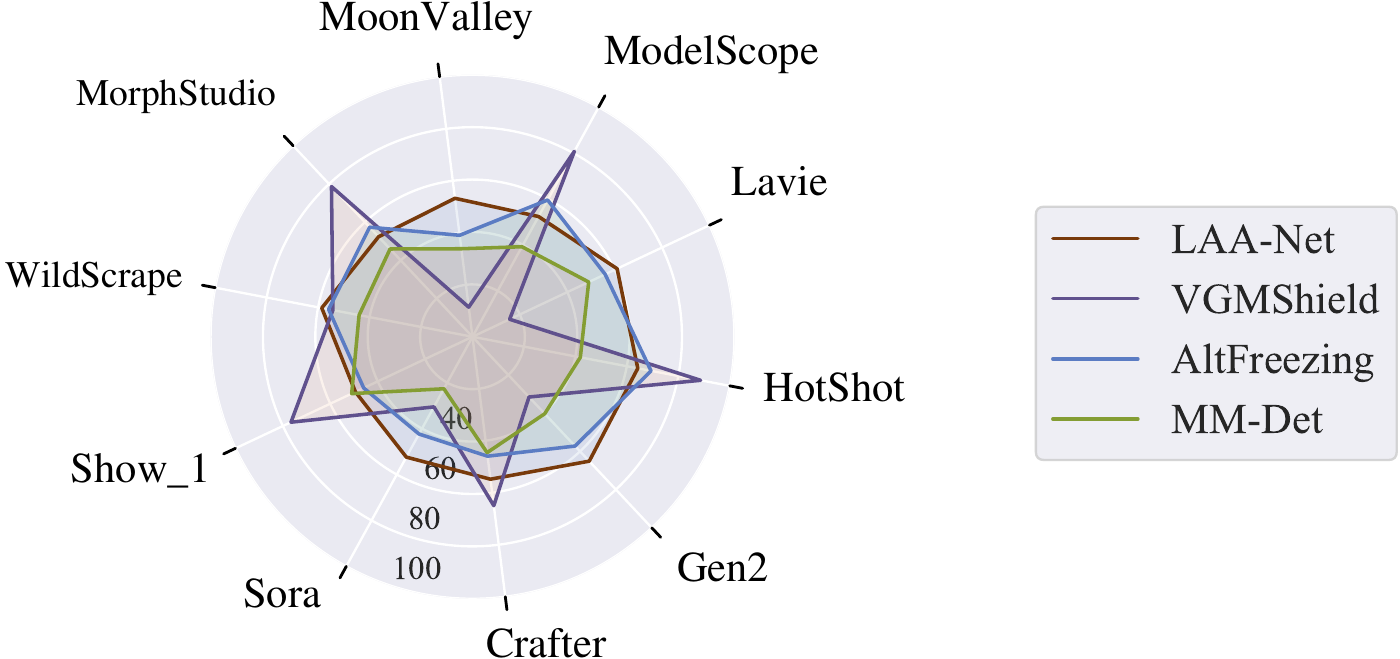}
        \caption{Video-level Detectors.}
        \label{fig:genvideo_9_3_datasets_radar_sub1}
    \end{subfigure}

    \begin{subfigure}{0.9\linewidth}
        \centering
        \includegraphics[width=0.9\linewidth]{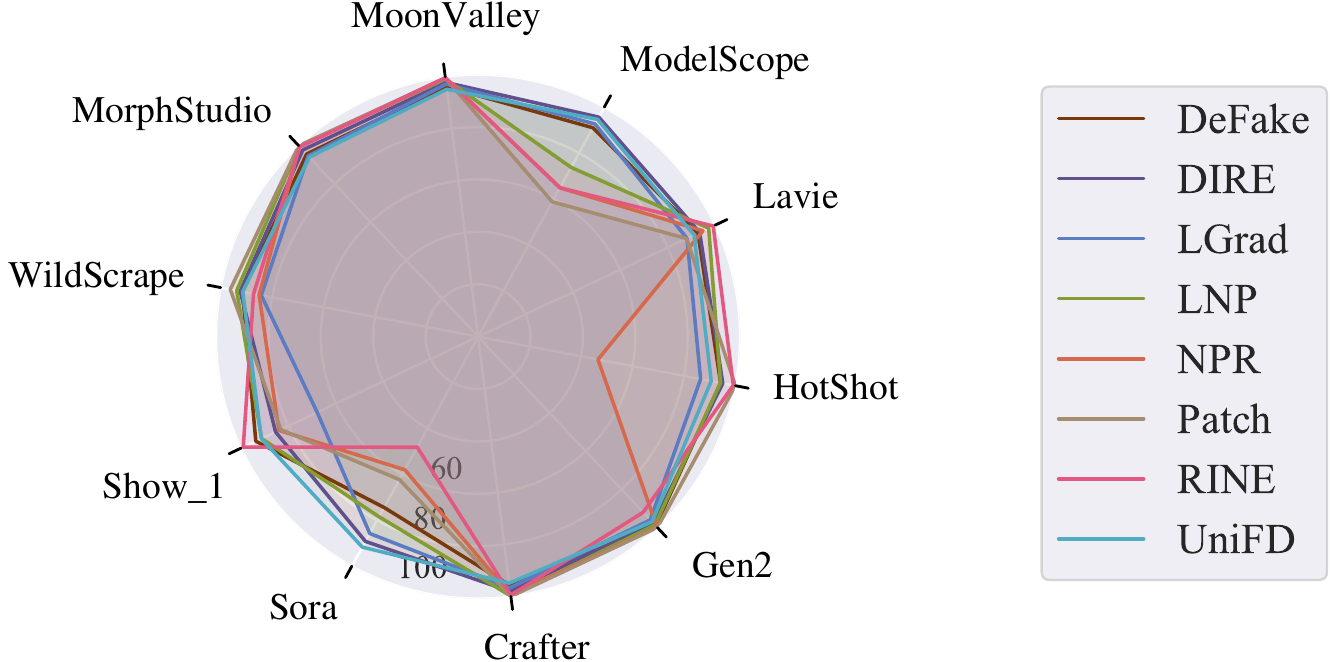}
        \caption{IV-Bridge Enhanced Detectors.}
        \label{fig:genvideo_9_3_datasets_radar_sub2}
    \end{subfigure}

    \caption{Results of cross-model detection AUC (\%) on GV Dataset (Part-II).}
    \label{fig:genvideo_9_3_datasets_radar}
\end{figure}

\begin{figure}[t]
    \centering

    \begin{subfigure}{0.9\linewidth}
        \centering
        \includegraphics[width=0.9\linewidth]{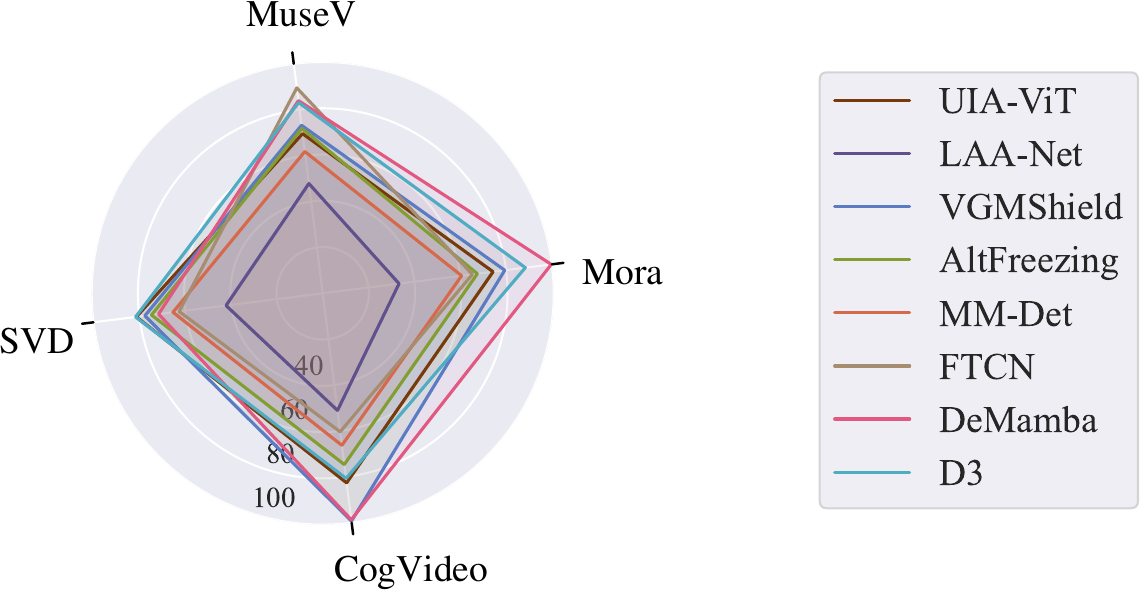}
        \caption{Video-level Detectors.}
        \label{fig:gvb_13_7_datasets_radar_sub1}
    \end{subfigure}

    \begin{subfigure}{0.9\linewidth}
        \centering
        \includegraphics[width=0.9\linewidth]{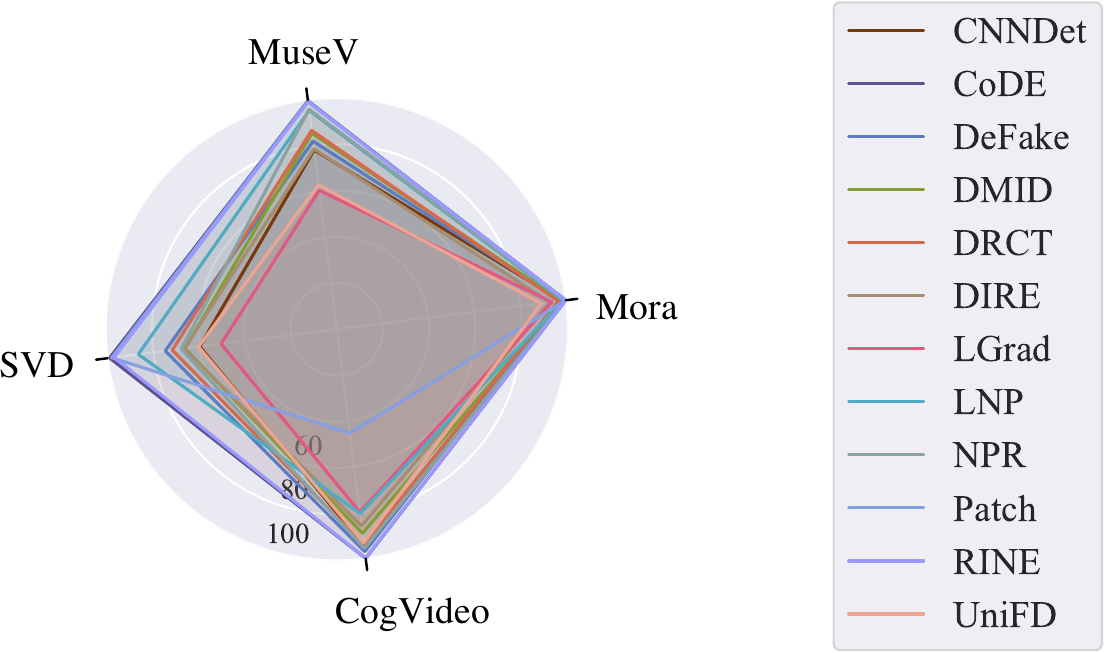}
        \caption{IV-Bridge Enhanced Detectors.}
        \label{fig:gvb_13_7_datasets_radar_sub2}
    \end{subfigure}

    \caption{Results of cross-model detection AUC (\%) on GVB Dataset.}
    \label{fig:gvb_13_7_datasets_radar}
\end{figure}

\begin{figure*}[t]
\centering
\includegraphics[width=1.0\textwidth]{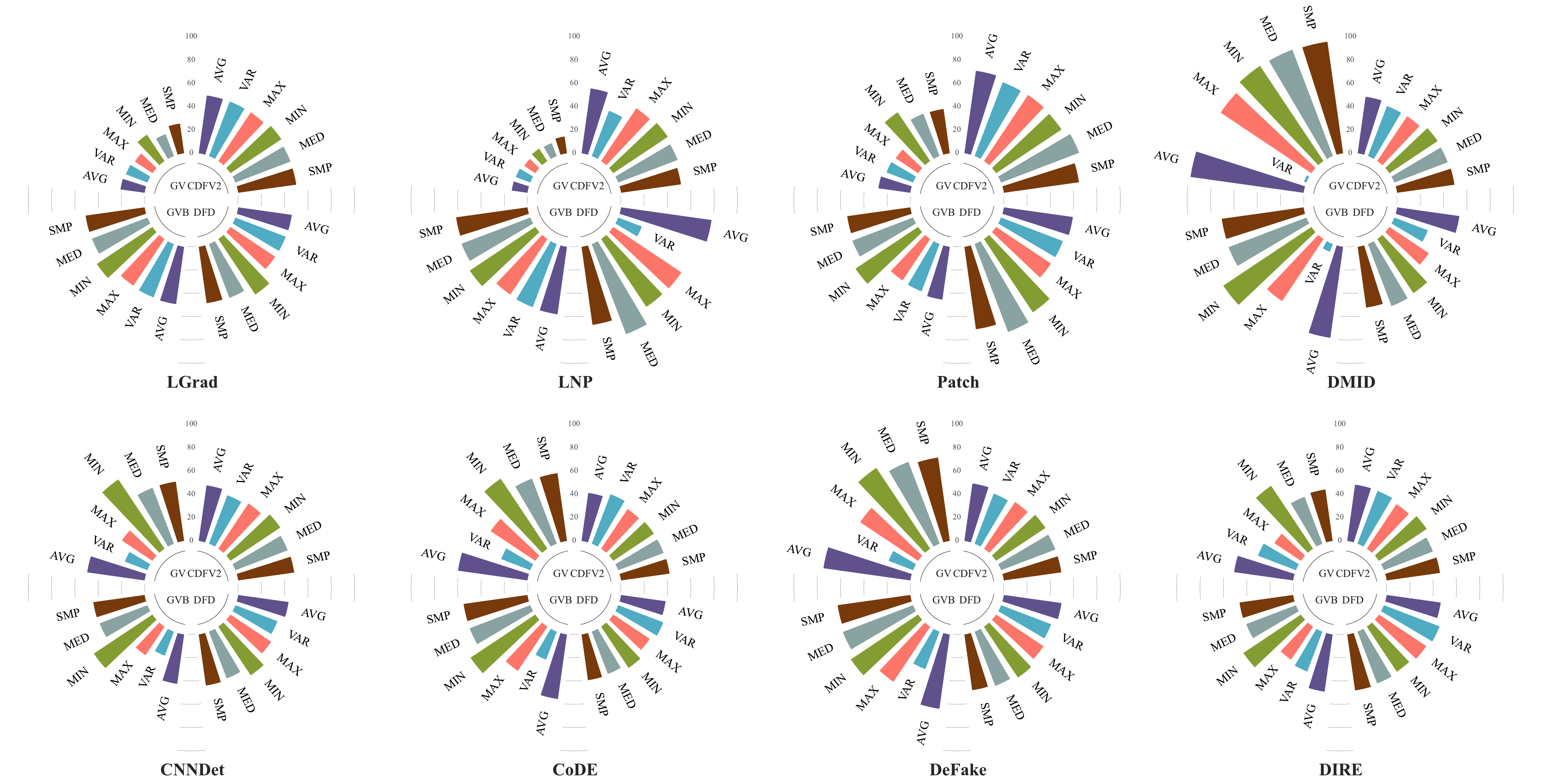}
\caption{The video-level performances of eight naive image-level detectors using six aggregation modes.}
\label{fig:8_of_12_video-level-mode-wo_vft}
\end{figure*}

\begin{table*}[t]
\caption{Overview of the video-level deepfake detectors included in \bench.}
\label{tab:video_level_deepfake_detection_methods}
\centering
\setlength{\tabcolsep}{10.5pt}
{
\renewcommand{\arraystretch}{0.9}
\begin{tabular}{l l l l l c l}
\toprule
\multirow{2}{*}{\textbf{Detector}} & 
\multirow{2}{*}{\textbf{Venue}} & 
\multirow{2}{*}{\textbf{Task}} & 
\multirow{2}{*}{\textbf{Architecture}} & 
\multirow{2}{*}{\textbf{Training Set}} &
\multicolumn{2}{c}{\textbf{vs. Image-level Detectors?}} \\
\cmidrule(lr){6-7}
& & & & & \textbf{Y/N} & \textbf{Method} \\
\midrule

FTCN~\cite{YJDMF21_1} & ICCV'21 & Facial & CNN + Transformer & FF++ & \yes & Average-level \\
UIA-ViT~\cite{WQZQHCZN22_1} & ECCV'22 & Facial & ViT & FF++ & \yes & Frame-level \\
AltFreezing~\cite{ZJWWH23_1} & CVPR'23 & Facial & 3DCNN & FF++  & \yes & Average-level \\
LAA-Net~\cite{DNPPMAED24_1} & CVPR'24 & Facial & EfficientNet-B4 + FPN & FF++ & \yes & Average-level \\ \midrule
VGMShield~\cite{YYT24_1} & ArXiv'24 & General & VideoMAE & TD  & \no & --\\
DeMamba~\cite{HYZZZYJHJWH24_1} & ArXiv'24 & General & XCLIP+Mamba & SEINE & \yes & Average-level \\
MM-Det~\cite{XXJQLXGX24_1} & NeurIPS'24 & General & VLLM+ViT & SVD & \yes & Average-level \\
D3~\cite{zheng2025d3} & ICCV'25 & General & XCLIP & - & \yes & Unknown \\
\bottomrule
\end{tabular}
}
\end{table*}

\begin{table*}[t]
\caption{Overview of the image-level deepfake detectors included in \bench.}
\label{tab:13_image_deepfake_detectors}
\centering
\setlength{\tabcolsep}{15.5pt}
{
\renewcommand{\arraystretch}{0.9}
\begin{tabular}{l l c c l c l}
\toprule

\multirow{2}{*}{\textbf{Detector}} & 
\multirow{2}{*}{\textbf{Venue}} & 
\multicolumn{2}{c}{\textbf{Training Set Domain}} &
\multirow{2}{*}{\textbf{Architecture}} & 
\multicolumn{2}{c}{\textbf{Learning-based?}} \\
\cmidrule(lr){3-4} \cmidrule(lr){6-7}
& & \textbf{GAN} & \textbf{Diffusion} & & \textbf{Y/N} & \textbf{Paradigm} \\
\midrule

CNNDet~\cite{SORAA20_1} & CVPR'20 & \yes & \no & ResNet & \yes & Data-driven \\ 
Patch~\cite{LDSP20_1} & ECCV'20 & \yes & \no & Xception & \yes & Data-driven \\ 
LNP~\cite{BFXBWX22_1} & ECCV'22 & \yes & \no  & ResNet & \yes & Specific Artifacts \\
LGrad~\cite{CYSGY23_1} & CVPR'23 & \yes & \no  & ResNet & \yes & Specific Artifacts \\ 
DeFake~\cite{ZZNY23_1} & CCS'23 & \no & \yes  & CLIP & \yes & Data-driven \\
DIRE~\cite{ZJWWHHH23_1} & ICCV'23 & \no & \yes & ResNet & \yes & Specific Artifacts \\
DMID~\cite{RDGGKL23_1} & ICASSP'23 & \no & \yes & ResNet & \yes & Data-driven \\
CoDE~\cite{LFMLAR24_1} & ECCV'24 & \no & \yes & ViT & \yes & Data-driven \\
DRCT~\cite{chen2024drct} & ICML'24 & \no & \yes & Conv-B/CLIP & \yes & Data-driven \\
UniFD~\cite{UYJ23_1} & CVPR'23 & \yes & \no  & CLIP & \yes & Data-driven \\
NPR~\cite{CYSGPY24_1} & CVPR'24 & \yes & \no  & ResNet & \yes & Specific Artifacts \\
RINE~\cite{CS24_1} & ECCV'24 & \yes & \no & ViT & \yes & Data-driven \\

\bottomrule
\end{tabular}
}
\end{table*}

\end{document}